\overfullrule=0pt
\input harvmac
\def\a{{\alpha}}

\def\l{{\lambda}}

\def\b{{\beta}}

\def\g{{\gamma}}

\def\om{{\omega}}
\def\d{{\delta}}
\def\e{{\epsilon}}

\def\half{{1\over 2}}
\def\p{{\partial}}

\def\t{{\theta}}
\def\tb{{\overline\theta}}
\def\bar{\overline}

\Title{\vbox{\hbox{IFT-P.011/2005 }}}
{\vbox{
\centerline{\bf Super-Poincar\'e Covariant Two-Loop Superstring Amplitudes}}}
\bigskip\centerline{Nathan Berkovits\foot{e-mail: nberkovi@ift.unesp.br}}
\bigskip
\centerline{\it Instituto de F\'\i sica Te\'orica, Universidade Estadual
Paulista}
\centerline{\it Rua Pamplona 145, 01405-900, S\~ao Paulo, SP, Brasil}

\vskip .3in
The super-Poincar\'e covariant formalism for the superstring is used to compute
massless four-point two-loop amplitudes in ten-dimensional superspace.
The computations are much simpler than in the RNS formalism and
include both external bosons and fermions.

\vskip .3in

\Date {March 2005}

\newsec{Introduction}

The computation of superstring multiloop amplitudes is important
for testing duality conjectures and verifying finiteness. In the RNS
formalism, the need to sum over spin structures leads to various
complications in multiloop computations. For 
two-loop amplitudes with four massless external Neveu-Schwarz states,
these complications were recently overcome in a series of papers by
D'Hoker and Phong
\ref\phong{E. D'Hoker and D.H. Phong, {\it Two Loop Superstrings, 
1. Main Formulas}, Phys. Lett. B529 (2002)
241, hep-th/0110247\semi
E. D'Hoker and D.H. Phong,
{\it Two Loop Superstrings, 2. The Chiral Measure on Moduli Space}, Nucl.
Phys. B636 (2002) 3, hep-th/0110283\semi
E. D'Hoker and D.H. Phong,
{\it Two Loop Superstrings, 3. Slice Independence and Absence
of Ambiguities}, Nucl. Phys. B636 (2002) 61, hep-th/0111016\semi
E. D'Hoker and D.H. Phong,
{\it Two Loop Superstrings, 4. The Cosmological Constant and Modular
Forms},
Nucl. Phys. B639 (2002) 129, hep-th/0111040\semi
E. D'Hoker and D.H. Phong,
{\it Two Loop Superstrings, 5. Gauge-Slice Independendence of the N-Point
Function}, hep-th/0501196\semi
E. D'Hoker and D.H. Phong,
{\it Two Loop Superstrings, 6. Non-Renormalization Theorems and the
Four-Point Function}, hep-th/0501197.},
and related papers by Iengo and Zhu et al \ref\zhuold{
R. Iengo and C.-J. Zhu, {\it
Two Loop Computation of the Four-Particle
Amplitude in the Heterotic String}, Phys. Lett. B212 (1988) 313\semi
R. Iengo and C.-J. Zhu, {\it Explicit Modular Invariant Two-Loop Superstring
Amplitude Relevant for $R^4$}, JHEP 06 (1999) 011, hep-th/9905050
\semi R. Iengo, {\it
Computing the $R^4$ Term at Two Superstring Loops}, JHEP 0202 (2002) 035,
hep-th/0202058\semi
Z.-J. Zheng, J.-B. Wu and C.-J. Zhu, {\it
Two-Loop Superstrings in Hyperelliptic Language I: the Main
Results}, Phys. Lett. B559 (2003) 89, hep-th/0212191\semi
Z.-J. Zheng, J.-B. Wu and C.-J. Zhu, {\it
Two-Loop Superstrings in Hyperelliptic Language II: the Vanishing
of the Cosmological Constant and the Non-Renormalization Theorem},
Nucl. Phys. B663 (2003) 79, hep-th/0212198\semi
Z.-J. Zheng, J.-B. Wu and C.-J. Zhu, {\it
Two-Loop Superstrings in Hyperelliptic Language III: the Four-Particle
Amplitude}, Nucl. Phys. B663 (2003) 95, hep-th/0212219\semi
W.-J. Bao and C.-J. Zhu, {\it Comments on Two-Loop Four-Particle
Amplitude in Superstring Theory}, JHEP 0305 (2003) 056, hep-th/0303152.}.
Since the final expression for the four-point two-loop amplitude is
remarkably simple, it is natural to ask if there is a more efficient
computational method than the RNS formalism.

Five years ago, a new super-Poincar\'e covariant formalism for the
superstring was introduced which uses pure spinors as worldsheet ghosts
\ref\superp{N. Berkovits, {\it Super-Poincar\'e Covariant Quantization of the
Superstring}, JHEP 04 (2000) 018, hep-th/0001035.}.
Last year, it was shown how to compute multiloop amplitudes using this
formalism and various vanishing theorems were proven \ref\loopold
{N. Berkovits, {\it Multiloop Amplitudes and Vanishing Theorems using
the Pure Spinor Formalism for the Superstring}, JHEP 0409 (2004) 047,
hep-th/0406055.}. In this paper,
this formalism will be used to compute massless four-point two-loop
amplitudes in ten-dimensional superspace. The computation is much simpler
than the RNS computations of \phong\zhuold\
and automatically includes both external
Neveu-Schwarz and external Ramond states.

As shown in \loopold, 
certain vanishing theorems related to finiteness are easily
proven in the super-Poincar\'e covariant formalism by counting fermionic
zero modes. To obtain the required number of fermionic zero modes for
massless multiloop amplitudes, one needs at least four external states.
And when there are precisely four external massless states, all fermionic
worldsheet variables contribute only through their zero modes. This
makes it relatively easy to evaluate four-point massless multiloop amplitudes
and the four-point two-loop amplitude will be explicitly computed here.
Higher-loop massless four-point amplitudes will hopefully
be discussed in a later paper.\foot{Zhu has recently proposed in
\ref\zhunew{C.-J. Zhu, {\it A Formula for Multiloop Four-Particle
Amplitude in Superstring Theory}, hep-th/0503001.}
a simple formula for four-point higher-loop Neveu-Schwarz
amplitudes which has many
features in common with the four-point two-loop Neveu-Schwarz expression.
Unfortunately, his proposal needs to be modified since his definition
of $\Delta_g(z_i, z_j)$ in equation (12) of the first
version of \zhunew\ is identically zero.
This follows from the fact that $\omega(P_I)=0$ for $I=1$ to $2g-2$,
implying that $\Delta_g(z_i,z_j)$ is proportional to
$\omega(z_i)\omega(z_j)$, which vanishes after antisymmetrizing in $i$ and $j$.
Nevertheless, preliminary computations using the super-Poincar\'e
covariant formalism
suggest that the polarization and momentum dependence of higher-loop
four-point amplitudes has the same structure as in the two-loop amplitude,
in agreement with the proposal of Zhu.}

\newsec{Four-Point Two-Loop Computation}

As discussed in \loopold, the four-point two-loop amplitude
for the Type IIB superstring\foot{This prescription is easily generalized
to the Type IIA or heterotic superstring.} is computed using the prescription:
\eqn\twoloop{{\cal A} = 
\int d^2\tau_1 d^2 \tau_2 d^2\tau_3~ \langle ~ |~
\prod_{P=1}^{3}\int d^2 u_P \mu_P(u_P)
\widetilde b_{B_P}(u_P,z_P) }
$$
\prod_{P=4}^{20} Z_{B_P}(z_P) \prod_{R=1}^{2} Z_J(v_R)
\prod_{I=1}^{11} Y_{C_I}(y_I)~|^2 ~\prod_{T=1}^4 \int d^2 t_T U^{(T)}(t_T)
~\rangle,$$
where $|~~|^2$ signifies the left-right product, 
$\tau_P$ are the Teichmuller parameters associated to the 
Beltrami differentials $\mu_P (u_P)$, $\widetilde b_{B_P}$
is the picture-raised $b$ ghost, $Z_{B_P}$ and $Z_J$ are the picture-raising
operators, $Y_{C_I}$ are the picture-lowering operators,
and $U^{(T)}(t_T)$ are the
dimension $(1,1)$ closed string
vertex operators for the four external states. 
Using the notation of \loopold, $\widetilde b_{B_P}$ satisfies
$\{Q, \widetilde b_{B_P}(u_P,z_P)\} = T(u_P) Z_{B_P}(z_P)$ where
\eqn\pictureop{Z_{B_P} = \half B^{mn} (\l\g_{mn} d) \d(B^{pq} N_{pq}),
\quad Z_J = \l^\a d_\a \d(J),\quad
Y_{C_I} = C_{I\a}\t^\a \d(C_{I\b}\l^\b),}
$N_{mn}$ and $J$ are the Lorentz and ghost-number currents for the 
pure spinors,
and $B_P^{mn}$ and $C_{I\a}$
are constant two-forms and spinors. As explained in \loopold, changing the 
choices for $B_P^{mn}$ and $C_{I\a}$ is a BRST-trivial operation
which does not affect the scattering amplitude.

For massless external states,
\eqn\formvc{
U^{(T)}= e^{ik\cdot x}(
\p \t^\a A_\a^{(T)} (\t) + 
\Pi^m A_m^{(T)} (\t) + d_\a W^{(T)\a} (\t) 
+\half N^{mn} {\cal F}_{mn}^{(T)}(\t)) }
$$ (\bar\p \tb^\b \bar A_\b^{(T)} (\tb) + 
\bar\Pi^p \bar A_p^{(T)} (\tb) + \bar d_\b \bar W^{(T)\b} (\tb) +
\half \bar N^{pq} 
\bar {\cal F}_{pq}^{(T)}(\tb))$$
where the Type IIB supergravity vertex operator has been written
as the left-right product of two super-Yang-Mills vertex operators.
Using the convention $D_\a = {\p\over{\p\t^\a}} + \half k_m (\g^m\t)_\a$,
\eqn\ymg{A_\a(\t) = \half a_m (\g^m\t)_\a -{1\over 3}
 (\xi\g_m\t)(\g^m\t)_\a + ... {\rm ~~~and
~~~}
A_m(\t) = a_m - (\xi\g^m\t) + ...}
are the spinor and vector gauge superfields and
\eqn\ymfs{W^\a(\t) = \xi^\a - {1\over 4}
k_{[m} a_{n]} (\g^{mn}\t)^\a + ... {\rm ~~~and
~~~}
{\cal F}_{mn} = k_{[m} a_{n]} - k_{[m} (\xi \g_{n]}\t) + ...}
are the spinor and vector superfield-strengths of super-Yang-Mills, where
$(a_m,\xi^\a)$ are the on-shell (gluon, gluino) and $...$ involves
$a_m$ and $\xi^\a$ with
higher powers of $k$ and $\t$.
The NS-NS, NS-R, R-NS, and R-R Type IIB supergravity vertex operators
can be obtained from \formvc\ by 
considering the terms
proportional to $a_m \bar a_n$, $a_m \bar \xi^\b$,
$\xi^\a \bar a_n$ and $\xi^\a \bar\xi^\b$ respectively.

At genus two, there are 16 fermionic zero modes for $\t^\a$ and 32
fermionic zero modes for $d_\a$. As in tree amplitudes, eleven of
the $\t^\a$ zero modes can come from the $Y_C$'s and the remaining
five $\t^\a$ zero modes will come from the external vertex operators. 
For the $d_\a$ zero modes, nineteen $d_\a$ zero modes can come from
the seventeen $Z_B$'s and two $Z_J$'s, so thirteen $d_\a$ zero modes
must come from the three $\widetilde b_B$ ghosts and
the four external vertex operators. 

{}From the construction of $\widetilde b_B$ in \loopold,
one finds that $\widetilde b_B$ contains terms with a maximum of four
$d$'s, but does not contain any terms with three $d$'s. Since each vertex
operator can contribute at most one $d_\a$ zero mode, the only contribution
from $\widetilde b_B$ comes from the terms with four $d$'s.
One can show that all such terms are proportional to
\eqn\proport{H_B(z)= B_{mn} B^{qr}~ (d(z)\g^{mnp} d(z))~(d(z)\g_{pqr}d(z)) 
~\d'(B^{st} N_{st}(z))}
where $\d'(x)$ denotes ${\p\over{\p x}}\d(x)$ and is defined to satisfy
$x \d'(x) = - \d(x)$.
Since each of the three $\widetilde b_B$ ghosts contains $\d'(BN)$
dependence, three of the vertex operators must contribute an $N_{mn}$ zero
mode to remove the derivative from the delta functions. Furthermore, the
fourth vertex operator must contribute the last of
the 32 $d_\a$ zero modes.

After performing the functional integration over the worldsheet nonzero modes,
the amplitude prescription of \twoloop\ gives
\eqn\twolooptwo{{\cal A} = 
\int d^2\tau_1 d^2 \tau_2 d^2\tau_3 \prod_{T=1}^4 ~\int d^2 t_T ~~
{{\exp(-\sum_{T,U=1}^4 k_T \cdot k_U G(t_T,t_U))}\over{(\det Im \Omega)^5}} }
$$
 |~
\int [DC][DB][D\l][D N]\int d^{16}\t d^{32} d $$
$$
\prod_{P=1}^{3}\int d^2 u_P \mu_P(u_P) H_{B_p}(u_P)
~\prod_{P=4}^{20} Z_{B_P}(z_P) \prod_{R=1}^{2} Z_J(v_R)
\prod_{I=1}^{11} Y_{C_I}(y_I) $$
$$\prod_{T=1}^4 (d_\a(t_T) W^{(T)\a}(\t) + \half N^{mn}(t_T) 
{\cal F}_{mn}^{(T)} (\t)) ~|^2 $$
where the factor of 
${{\exp(-\sum_{T,U=1}^4 k_T \cdot k_U G(t_T,t_U))}\over{(\det Im \Omega)^5}} $
comes from the functional integration over the ten $x$'s, $\Omega$ is the
period matrix,
$G(t_T,t_U)$ is the usual scalar Green's function, and
$\int [DC][DB][D\l][D N]$ are measure factors for the pure spinor
zero modes which are defined in \loopold. The partition function
vanishes in this formalism since the contribution from the
ten $x^m$ and 32 $(d_\a,\t^\a)$ variables
cancels the contribution from the 22 pure spinor variables.

To evaluate \twolooptwo, first use the rules described in \loopold\
to integrate
over the zero modes of $N_{mn}$ and $d_\a$ and over the choices of $B_{mn}$.
This produces the expression
\eqn\twoloopthree{{\cal A} = 
\int d^2\tau_1 d^2 \tau_2 d^2\tau_3 \prod_{T=1}^4 \int d^2 t_T
{{\exp(-\sum_{T,U=1}^4 k_T \cdot k_U G(t_T,t_U))}\over{(\det Im \Omega)^5}} }
$$
 |~
\int [DC][D\l]\int d^{16}\t  $$
$$
\prod_{P=1}^{3}\int d^2 u_P \mu_P(u_P) \Delta(u_1,u_2)\Delta(u_2,u_3)\Delta
(u_3,u_1)~ \prod_{I=1}^{11} Y_{C_I}(y_I) ~
 \l^\a\l^\b \l^\g (\g^{mnpqr})_{\a\b} \g^s_{\g\d} $$
$$
(~F_{mn}^{(1)}(\t)
F_{pq}^{(2)}(\t) F_{rs}^{(3)}(\t) W^{(4)\d}(\t) ~
\Delta(t_1,t_3)\Delta(t_2, t_4) + {\rm permutations ~~of ~~} 1234 ~) ~|^2$$
where $\Delta(u,v) = \e^{CD} \om_C(u) \om_D(v)$ and
$\omega_C(z)$ for $C=1,2$ are the two holomorphic one-forms.

To derive \twoloopthree\ from \twolooptwo, one uses that each
$H_{B_P}(u_P)$ has $+2$ conformal weight, has no poles on the surface,
and has zeros when $u_{P_1}=u_{P_2}$. The
unique such function is proportional to
$\Delta(u_1,u_2)\Delta(u_2,u_3)\Delta(u_3,u_1)$. Similarly, 
the picture-raising operators have zero conformal weight and no poles,
so they leave no contribution. And the external vertex
operators have $+1$ conformal weight with no poles, so they contribute
$h^{CDEF} \om_C(t_1)\om_D(t_2)\om_E(t_3)\om_F(t_4)$ for some constant
$h^{CDEF}$. Since the zero modes
associated with $\om_1$ and $\om_2$ appear symmetrically, 
$h^{CDEF}$ vanishes
unless it has two 1 indices and two 2 indices, and
is invariant under the exchange of the two 1 indices
with the two 2 indices. 

Moreover, Lorentz invariance implies that the three
remaining $\l$'s must be contracted with the indices of the external
superfields as
\eqn\unique{ \l^\a\l^\b \l^\g (\g^{mnpqr})_{\a\b} \g^s_{\g\d} 
~F_{mn}^{(1)}(\t) F_{pq}^{(2)}(\t) F_{rs}^{(3)}(\t) W^{(4)\d}(\t) }
up to permutations of the external superfields.\foot{
This contraction can be shown to be unique by decomposing the
(Wick-rotated) $SO(10)$ representations into $SU(5)\times U(1)$
representations. Under $SU(5)\times U(1)$, $W^\a$ decomposes into
$[W^+_{5\over 2}, W_{\half[ab]}, W_{-{3\over 2}}^a]$
and $F_{mn}$ decomposes into $[F_{+2}^{[ab]}, F_{0a}^b, F_{-2[ab]}]$
where $a,b=1$ to 5 and the subscript is the $U(1)$ charge. 
Choosing $\l^\a$ such that $\l^+_{5\over 2}$ is the only nonzero component,
one can easily verify that 
\eqn\uniquetwo{ (\l^+_{5\over 2})^3 F_{-2[ab]}^{(1)}
F_{-2[cd]}^{(2)}
F_{-2[ef]}^{(3)} W^{(4)f}_{-{3\over 2}} \e^{abcde}}
is the unique $SU(5)\times U(1)$ invariant term, 
which is written in $SO(10)$-invariant notation as \unique.}
And 
$\l^\a\l^\b \l^\g (\g^{mnpqr})_{\a\b} \g^s_{\g\d} 
(F_{mn}^{(1)} F_{pq}^{(2)} F_{rs}^{(3)} +
F_{mn}^{(2)} F_{pq}^{(3)} F_{rs}^{(1)} +
F_{mn}^{(3)} F_{pq}^{(1)} F_{rs}^{(2)} )=0$ together with
$(\g^{mnpqr})_{\a\b}  
(F_{mn}^{(1)} F_{pq}^{(2)} - 
F_{mn}^{(2)} F_{pq}^{(1)})  
=0$ 
implies
that one can replace $h^{CDEF}$ with $\e^{CE}\e^{DF}$.
Note that by choosing the Teichmuller parameters to 
be the three elements of the period matrix $\Omega_{CD}$, 
one can write
$$\int
d^2\tau_1 d^2 \tau_2 d^2\tau_3 
|\prod_{P=1}^{3}\int d^2 u_P \mu_P(u_P) \Delta(u_1,u_2)\Delta(u_2,u_3)\Delta
(u_3,u_1)|^2 = 
\int d^2 \Omega_{11} d^2 \Omega_{12} d^2 \Omega_{22}.$$

Finally, the integration over $\int [DC][D\l]\int d^{16}\t$
in \twoloopthree\ is easily performed using the rules of \loopold\ to obtain
\eqn\twoloopfour{{\cal A} = 
\int d^2\Omega_{11} d^2 \Omega_{12}  d^2 \Omega_{22}
\prod_{T=1}^4 \int d^2 t_T
{{\exp(-\sum_{T,U=1}^4 k_T \cdot k_U G(t_T,t_U))}\over{(\det Im \Omega)^5}} }
$$
 |~
 (\g^{mnpqr})_{\a\b} \g^s_{\g\d} ~~(\int d^5\t)^{\a\b\g} $$
$$
(~F_{mn}^{(1)}(\t)
F_{pq}^{(2)}(\t) F_{rs}^{(3)}(\t) W^{(4)\d}(\t) 
~\Delta(t_1,t_3)\Delta(t_2, t_4) + {\rm permutations ~~of ~~} 1234 ~) ~|^2 $$
where 
\eqn\defint{(\int d^5\t)^{\a\b\g} =
 (T^{-1})^{\a\b\g}_{\rho_1 ...\rho_{11}}
\e^{\rho_1 ...\rho_{16}} ({\p\over{\p\t}})_{\rho_{12}} ...
({\p\over{\p\t}})_{\rho_{16}} }
and $
(T^{-1})^{\a\b\g}_{\rho_1 ...\rho_{11}}$ is the $\g$-matrix traceless
part of 
$$\e_{\rho_1 ...\rho_{16}}(\g^m)^{\a\rho_{12}}
(\g^n)^{\b\rho_{13}}
(\g^p)^{\g\rho_{14}}
(\g_{mnp})^{\rho_{15}\rho_{16}}.$$
In other words,
\eqn\deftinv{(T^{-1})^{\a\b\g}_{\rho_1 ...\rho_{11}} =
\e_{\rho_1 ...\rho_{16}}(\g^m)^{\a\rho_{12}}
(\g^n)^{\b\rho_{13}}
(\g^p)^{\g\rho_{14}}
(\g_{mnp})^{\rho_{15}\rho_{16}} + \g_m^{(\a\b} 
E^{\g) m}_{\rho_1 ...\rho_{11}}}
where $ E^{\g m}_{\rho_1 ...\rho_{11}}$ is defined such that
$\g^m_{\a\b} (T^{-1})^{\a\b\g}_{\rho_1 ...\rho_{11}}=0$.

The four-point two-loop amplitude of  \twoloopfour\ is remarkably
simple. When all external states are chosen in the NS-NS sector, it
should not be difficult to show agreement with the RNS formula of \phong\zhuold.
Work is currently
in progress on extending these results to higher-loop four-point
amplitudes.

\vskip 15pt
{\bf Acknowledgements:} I would like to thank  
Chuan-Jie Zhu for useful discussions,
CNPq grant 300256/94-9, 
Pronex 66.2002/1998-9,
and FAPESP grant 04/11426-0
for partial financial support, and the ICTP 
and Funda\c{c}\~ao Instituto de F\'{\i}sica Te\'orica  
for their hospitality.

\listrefs

\end